\documentclass[10pt,prl,twocolumn,nofootinbib,byrevtex,a4paper]{revtex4}
\usepackage{microtype}
\usepackage{graphics}
\usepackage{amssymb}

\newcommand{\be}{\begin{equation}}
\newcommand{\ee}{\end{equation}}
\newcommand{\ra}{\rightarrow}

\begin{document}

\title{Comment on ``Towards a large deviation theory for strongly correlated systems''}

\author{Hugo Touchette}
\email{ht@maths.qmul.ac.uk}
\affiliation{School of Mathematical Sciences, Queen Mary University of London, London E1 4NS, UK}

\date{\today}

\begin{abstract}
I comment on a recent paper by Ruiz and Tsallis [Phys.\ Lett.\ A \textbf{376}, 2451 (2012)] claiming to have found a ``$q$-exponential'' generalization of the large deviation principle for strongly correlated random variables. I show that the basic scaling results that they find numerically can be reproduced with a simple example involving independent random variables, and are not specifically related to the $q$-exponential function. In fact, identical scaling results can be obtained with any other power-law deformations of the exponential. Thus their results do not conclusively support their claim of a $q$-exponential generalization of the large deviation principle.
\end{abstract}

\maketitle

Ruiz and Tsallis have published a recent a paper \cite{ruiz2012} in which they claim to have found a generalized large deviation approximation, or large deviation principle (LDP),\footnote{Following the standard terminology of large deviation theory, I use the term `large deviation principle' to mean `large deviation approximation' \cite{touchette2009}.} for a process involving strongly correlated random variables. This LDP supposedly takes the form not of an exponential, as in standard large deviation theory \cite{touchette2009}, but of a so-called $q$-exponential function, which is one possible generalization of the exponential function. My purpose here is to show that the main results of \cite{ruiz2012}, which take the form of scaling plots, are not specifically related to $q$-exponentials or correlated random variables, and do not conclusively point, as claimed, to a $q$-exponential generalization of the LDP and large deviation theory. I illustrate these points with a simple example that shows how one might be misled into defining ``$q$-exponential LDPs'' for distributions that do not have in fact the form of a $q$-exponential. 

The example is a classic one in large deviation theory showing that not all sums of random variables have an LDP \cite{lanford1973,touchette2009}. Consider the sample mean
\be
S_n=\frac{1}{n^\beta}\sum_{i=1}^n X_i,
\label{eqsum1}
\ee
and assume that the $X_i$'s are independent random variables identically distributed according to the Cauchy distribution:
\be
p(X_i=x)=\frac{1}{\pi}\frac{\gamma}{x^2+\gamma^2},
\ee
where $\gamma$ is a positive parameter that controls the width of the distribution. At this point I should say that, since the $X_i$'s are assumed to be mutually independent, one might be tempted to discard this example as being unrelated to the results of Ruiz and Tsallis \cite{ruiz2012}. However, it will become clear that the properties of $S_n$ are in fact relevant to discuss their results.

Since Cauchy random variables are stable \cite{gnedenko1954}, the probability density of $S_n$ is also a Cauchy distribution:
\be
p(S_n=s)=\frac{1}{\pi} \frac{\gamma'}{s^2+\gamma'^2},\qquad \gamma'=\frac{\gamma}{n^{\beta-1}}.
\label{eqpdf1}
\ee 
With this result we can see that $p(S_n=s)$ concentrates to the Dirac-delta function $\delta(s)$ in the limit $n\ra\infty$ whenever $\beta>1$. Hence the density of $S_n$ concentrates in some limit, which is the first ingredient that we need to define an LDP. However, since
\be
p(S_n=s)\sim \frac{\gamma}{n^{\beta-1}s^2}
\label{eqs1}
\ee
as $n\ra\infty$, the concentration to the delta function is not exponential in $n$, which implies that $p(S_n=s)$ does not satisfy an LDP. In fact, if we try to define an LDP for $S_n$ using the standard large deviation limit,
\be
I(s)=\lim_{n\ra\infty} -\frac{1}{n}\ln p(S_n=s),
\ee
we find that the rate function $I(s)$ is $0$ for all $s$. This means that $S_n$ does not satisfy an LDP. One could say that it has a zero rate function, but this gives us no information about $S_n$ apart from knowing that the concentration of $p(S_n)$ is slower than exponential. The same conclusion applies to any sums of (symmetric) L\'evy-stable random variables.

From this it is tempting to change the definition of the LDP in the following way. Since we know that $p(S_n=s)$ decays to zero for $s\neq 0$ as a polynomial in $n$, we could try to quantify this decay by a rate function defined on a scale different from the logarithm. This is in essence the proposal of Ruiz and Tsallis \cite{ruiz2012}. In particular, we could replace, as they do, the normal log with the so-called $q$-logarithm function defined by
\be
\ln_q(x)=\frac{x^{1-q}-1}{1-q},
\ee
which converges to the normal logarithm for $q\ra 1$. 

\begin{figure}[t]
\includegraphics{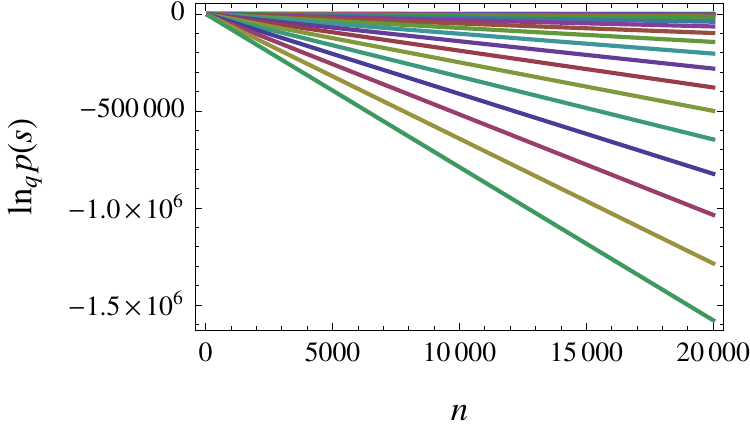}
\caption{Plot of the $q$-log of the probability density $p(s)$ of $S_n$ for $\beta=1.5$ as a function of $n$. Different lines correspond to different values of $s$ (from 0 to 2 in steps of 0.1, from top to bottom). The plot is for $q=3$, the only value of $q$ for which this plot yields straight lines.}
\label{figqscale1}
\end{figure}

For the Cauchy example, it is easy to see that
\be
\ln _q p(S_n=s)\sim \frac{n^{(1-\beta)(1-q)}}{1-q}
\ee
as $n\ra\infty$, so we can choose
\be
q=\frac{\beta}{\beta-1}>1
\label{eqq1}
\ee 
to obtain
\be
\ln_q p(S_n=s)\sim -c_s n,
\label{eqqs1}
\ee
where $c_s$ is some constant that depends only on $s$. This behavior is illustrated in Fig.~\ref{figqscale1} with $\beta=1.5$ for which $q=3$. Any value of $q$ not given by Eq.~(\ref{eqq1}) will yield a plot of $\ln_q p(S_n=s)$ that is not linear in $n$, as can easily be checked analytically and numerically.

If we are to follow Ruiz and Tsallis \cite{ruiz2012}, then the straight lines seen on the $q$-log plot of $p(S_n=s)$ should imply that this density has the form of a $q$-exponential,
\be
e_q^x=[1+(1-q)x]^{\frac{1}{1-q}},
\ee 
the inverse of the $q$-logarithm. But, as can be seen from Eq.~(\ref{eqpdf1}), this is an incorrect conclusion: $p(S_n=s)$ does not have the form of a $q$-exponential, although it has the same tails as this function, so the straight lines that we see are not a conclusive sign of a $q$-exponential LDP---they are only an indication of power-law tail behavior. Any density that has power-law tails will lead to straight lines over a wide range of values when plotted on a $q$-log scale, and this even if its center does not behave as a power-law, as is the case for the Cauchy sum. Moreover, it will show straight lines when plotted with any other power-law deformations of the exponential, so there is nothing special about the $q$-log and the $q$-exp.

\begin{figure}[t]
\includegraphics{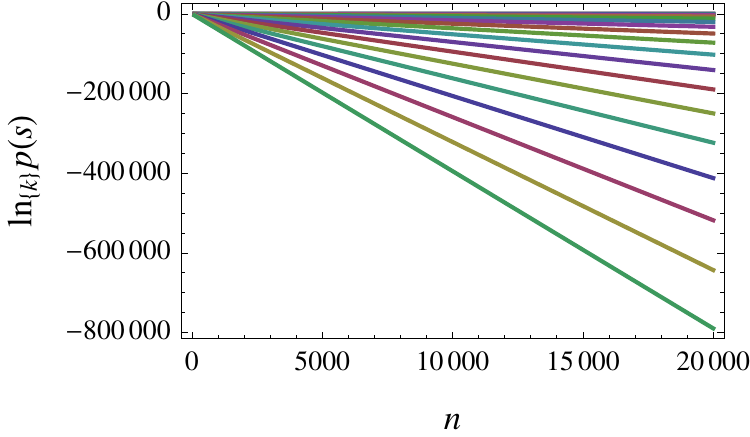}
\caption{Plot of the $\kappa$-log of the probability density $p(s)$ of $S_n$ for $\beta=1.5$ as a function of $n$ for different values of $s$ (from 0 to 2 in steps of 0.1, from top to bottom). The plot is for $\kappa=2$, the only value of $\kappa$ for which this plot yields straight lines.}
\label{figkscale1}
\end{figure}

To illustrate this point, consider the so-called $\kappa$-logarithm introduced by Kaniadakis \cite{kaniadakis2001},
\be
\ln_{\{\kappa\}} x=\frac{x^{-\kappa}-x^{\kappa}}{2\kappa},
\ee
which recovers the normal log as $\kappa\ra 0$. This deformation of the log is obviously different from the $q$-log; yet it can be used in the same way as before to transform $p(S_n=s)$ into a seemingly linear function of $n$. In this case, we have to choose $\kappa=q-1$ to obtain
\be
\ln_{\{\kappa\}} p(S_n=s)\sim -d_s n
\ee
as $n\ra\infty$. This is shown in Fig.~\ref{figkscale1}. 

From this it is clear that if Ruiz and Tsallis \cite{ruiz2012} had decided to use the $\kappa$-log instead of the $q$-log, they would have obtained similar results, i.e., straight lines that are neither necessarily related to the $q$-exp nor the $\kappa$-exp, the inverse of the $\kappa$-log. Therefore, the main result found in their abstract and their Eq.~(16), which consists of an LDP expressed in terms of the $q$-exp, is misleading: the only conclusion that follows from their plots is that they study a distribution with power-law tails, having the same asymptotic form as any power-law deformation of the exponential. 

Of course, the reason why the $\kappa$-log does the same job as the $q$-log is that they both behave as inverse power-laws for small arguments, so that a function with power-law tails must appear linear when transformed with either function. From this simple reasoning, it is clear that any other power-law deformation of the logarithm (there is an infinite number of them) will do the job. In fact, we do not even need to consider deformations of the logarithm: by inserting $p(S_n=s)$ into
\be
f_r(x)=\frac{x^r}{r},
\ee
we can choose $r=1/(1-\beta)$ to obtain 
\be
f_r(p(S_n=s))\sim - n,
\ee
which is just a different way of expressing the power-law behavior of $p(S_n=s)$. The same trick can be applied to the distribution considered in \cite{ruiz2012} with the same result.

In the end, one could try to distinguish whether the distribution of a random variable such as $S_n$ \emph{exactly} follows a $q$-exp, a $\kappa$-exp or another power-law deformation of the exponential by studying its behavior for small values of $n$. However, this would defeat the purpose of large deviation theory, which is to derive general asymptotic results by considering specific scaling limits of stochastic processes, such as the large-$n$ limit studied here, or the long-time or low-noise limits \cite{touchette2009}. It is by considering these limits that general asymptotics are seen to arise in a broad class of stochastic processes.

The results of \cite{ruiz2012} can be taken as a sign of an interesting power-law scaling in the tails of a probability distribution. But, as I have shown, they are insufficient to conclude that this scaling has the form of a $q$-exponential or that it is specifically related to correlated random variables, as implied in \cite{ruiz2012}. Moreover, as these results only concern one specific process, they are obviously not enough to conclude that one has generalized large deviation theory to strongly correlated random variables. As of now, no such generalization exists.

\bibliography{qldt}

\end{document}